# Quasi-static and propagating modes in three-dimensional THz circuits

Mathieu Jeannin,[1] Djamal Gacemi,[1] Angela Vasanelli,[1] Lianhe Li,[2] Alexander Giles Davies,[2] Edmund Linfield,[2] Giorgio Biasiol,[3] Carlo Sirtori[1] and Yanko Todorov[1,*]

[1]*Laboratoire de Physique de l'École Normale Supérieure, ENS, Université PSL, CNRS, Sorbonne Université, université de Paris, F-75005 Paris, France*

[2]*School of Electronic and Electrical Engineering, University of Leeds, LS2 9JT Leeds, United Kingdom*

[3]*Laboratori TASC, CNR-IOM at Area Science Park, Strada Statale 14, km163.5, Basovizza, TS34149, Italy*

*yanko.todorov@ens.fr

**Abstract:** We provide an analysis of the electromagnetic modes of three-dimensional metamaterial resonators in the THz frequency range. The fundamental resonance of the structures is fully described by an analytical circuit model, which not only reproduces the resonant frequencies but also the coupling of the metamaterial with an incident THz radiation. We also evidence the contribution of the propagation effects, and show how they can be reduced by design. In the optimized design the electric field energy is lumped into ultra-subwavelength ($\lambda/100$) capacitors, where we insert semiconductor absorber based on the collective electronic excitation in a two dimensional electron gas. The optimized electric field confinement is evidenced by the observation of the ultra-strong light-matter coupling regime, and opens many possible applications for these structures for detectors, modulators and sources of THz radiation.

## 1. Introduction

The majority of metamaterials has a two-dimensional geometry and benefit from well-established planar top-down fabrication techniques. When used as passive optical components, this two-dimensional character results in flat optical elements, such as lenses or phase control/phase shaping devices [1]–[4]. Bottom-up fabrication techniques involving dielectric material deposition can be used to create layered metamaterials consisting in stacked, two-dimensional meta-layers [5]–[9]. However, this approach introduces strong limitations when a semiconductor active region has to be employed, e.g. in the case of active components such as emitters, detectors and modulators. Indeed, in planar structures the interaction with the active region only relies on the evanescent field that decays away from the two-dimensional metamaterial [10]–[12]. An alternative solution resides in double metal geometries embedding the active layer between two metal planes using Au-Au wafer bonding [13]–[17]. While it is well-established in the mid-IR and THz domains, where the losses associated to the metallic confinement are moderate, a fabrication constraint of this technique is the fact that the bottom metallic plane cannot be shaped, thus strongly limiting the possible designs, especially when injection or extraction of electrical current into the active region of the device is required. Recently, a few demonstrations of truly three-dimensional architectures have been published, where the metallic ground plane can be partly etched [18], or even replaced by a lithographically designed pattern [19], [20]. In refs. [19]–[21] we have proposed a three-dimensional THz LC architecture sustaining a circuit-like mode where the electric field is strongly confined inside the capacitive parts of the device that are filled with a semiconductor layer, and a similar design was proposed for THz modulators [22]. A key property targeted in such double-metal designs is their capacity to confine the electric field of the resonant mode into well-defined regions of space that correspond exactly to their capacitive parts. This can be achieved when the fundamental mode of the structure is truly quasi-static, i.e. when propagation effects are minimized. While this regime is common in the low-frequency part of the electromagnetic spectrum [23]–[25], it is more challenging at THz and higher frequencies. Indeed, according to Maxwell's equations, propagation effects are driven by the displacement current term $\varepsilon_0 d\mathbf{E}/dt = -i\varepsilon_0 \omega \mathbf{E}$, (where $\mathbf{E}$ is the electric field of the resonator and $c$ is the speed of light) which becomes progressively more important as the resonance frequency $\omega$ is increased [26].

In this paper, we provide a comprehensive analysis on the different modes of our metamaterial device, evidencing the presence of such quasi-static, circuit-like mode. In the case of the fundamental resonance, we discuss in detail the optimization of the effective mode volume, which is limited by either propagation effects or fringing field effects. Propagation effects are discussed within an analytical circuit model, which also allows quantifying the coupling of these structures with incident THz radiation. The model explicitly treats different parts of the resonator as lumped circuit elements, and can be used when combining such structures with different active media in applications. Indeed, it is always possible to map any given electromagnetic resonance into an effective LC resonator model. However, at THz frequencies and higher it becomes increasingly difficult to construct a truly lumped element circuit because of the aforementioned contribution of the displacement current. In our structure, the resonance frequency and the response to the incident THz radiation are solely a function of the lumped circuit geometry that define the lowest order LC resonance. These developments have led us to an optimized design of the THz circuit, which was recently used for studies of the ultra-strong light-matter coupling regime [20].

## 2. Resonator fabrication and characterization

Our metamaterial resonator is presented in Fig. 1(a). It is based on the optimization of the top hat resonator (THR) of Ref. [16]. It consists of a bottom metal strip (dashed black rectangle in the SEM image) of fixed width (1 µm) and length (3 µm). The top metal pattern serves as a square inductive loop, with a variable arm length $Y$. The top and bottom parts overlap vertically, defining two 1 µm x 1 µm parallel plate capacitors (red thin lines in the SEM image). The arms of the top pattern are terminated by small extensions that allow a better coupling to free-space radiation [27]. A 1.33µm thick semiconductor rod is contained between the bottom and top metal patterns (light gray area in the SEM image), and the whole meta-atom is embedded in a SiN layer of thickness $T_{SiN}$ = 3 µm, above a metallic mirror. The semiconductor active region comprises 20 repetitions of 32 nm thick GaAs quantum wells (QWs) separated by 20 nm $Al_{0.15}Ga_{0.85}As$ barriers. The QWs are modulation doped by Si delta-doped regions placed 5 nm away from the QW, with a nominal sheet carrier density of 2 x $10^{11}$ cm$^{-2}$. The fabrication procedure has been described elsewhere [20].

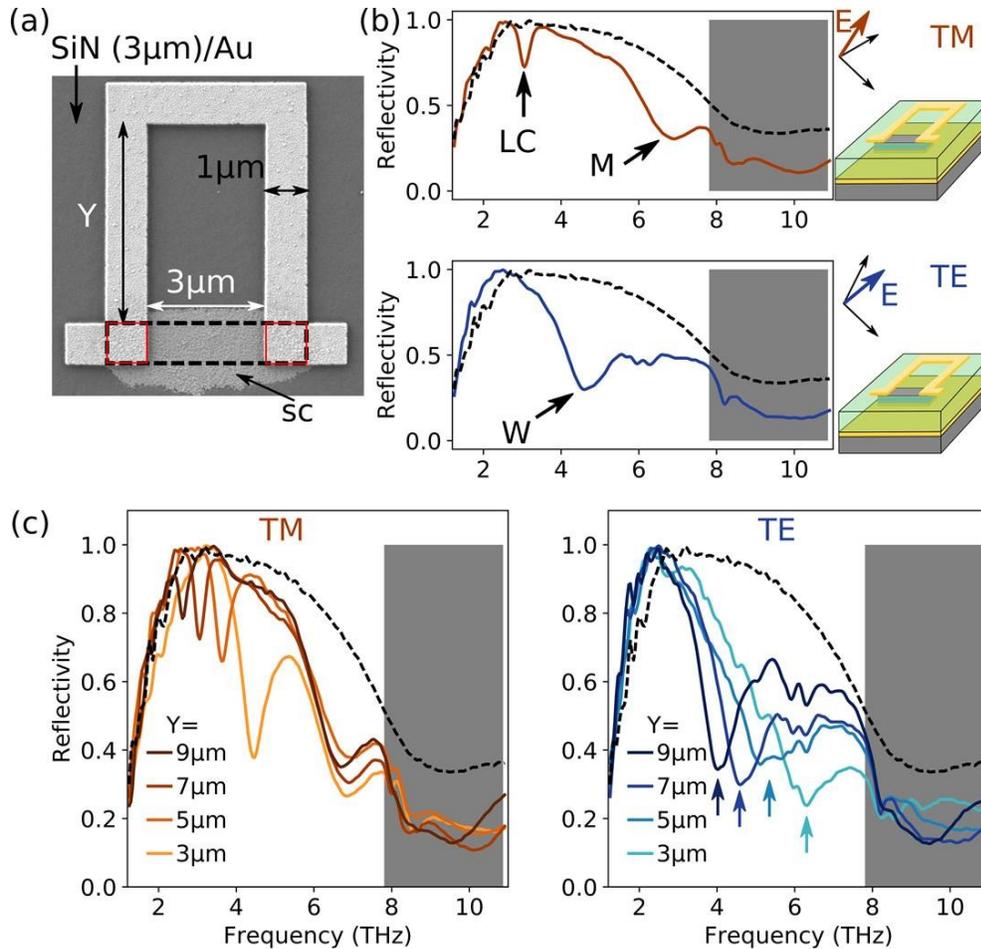

Figure 1: *(a) Scanning electron microscope image of a THz LC circuit, showing the top Au inductive loop. The bottom plate is delimited by the dashed rectangle. The capacitor area is shown in thin red lines. The dark grey background corresponds to 3µm of SiN on top of a gold mirror, while the light grey rod is the semiconductor (SC) active region. (b) Typical reflectivity spectrum of an array of THz LC circuits for the TM (top) and TE (bottom) polarizations, as depicted in the insets on the right. We see two resonances, indicated as LC and M for the TM polarization, and only one resonance (indicated as W) for the TE polarization. . The reflectivity of the layered substrate is shown in by the black dashed lines. (c) Reflectivity spectra for different inductor length Y (indicated in the legend, in µm).*

We fabricate arrays of ~6000 resonators covering a surface of ~2 x 2 mm$^2$ with different inductor lengths Y = 3, 5, 7, 9 µm. We probe their optical properties in a reflectivity experiment using linearly polarized light from the globar source of a Fourier transform infrared (FTIR) spectrometer (Bruker 70v). The experiment is performed at room temperature, where the electronic population of the first two confined subbands is almost equal, such that the coupling of the resonator mode and the electronic transitions in the quantum wells can be neglected. Light impinges on the sample at an angle of 15°, and the electric field is either TM or TE polarized, as depicted in the insets of Fig. 1(b). All spectra are normalized to the reflectivity of a flat gold surface. We use a He-cooled Ge bolometer (QMC instruments) as a detector. A typical spectrum for each polarization is shown in Fig. 1(b) along with the reflectivity from the bare SiN/Au layered substrate (black dashed line). In the case of TM-polarized light, the projection of the electric field in the sample plane is along the line defined by the two capacitors, and we observe two well-isolated absorption features, marked as LC and M, as well as the presence of the GaAs Reststrahlen band arising from the optical phonons (grey area). In the case of the TE-polarized light, the electric field is oriented along the variable length arm of the THR. In this polarization the reflectivity spectrum shows only one pronounced absorption feature labelled W, and the GaAs Reststrahlen band. In Fig. 1(c) we show the reflectivity spectra for different inductor lengths. The TM-polarized spectra (left panel) demonstrate that the LC mode red-shifts with increasing inductor length, while its contrast decreases. On the other hand, the M mode is barely affected by the change in inductor length. The TE-polarized spectra (right panel) show that the mode W also red-shifts with increasing inductor length (blue arrows).

To understand the nature of these resonances better, we performed numerical simulations of the metamaterial response using commercial finite element method software (Comsol Multiphysics v5.2). The results of these simulations are presented in Figure 2. We show the calculated reflectivity spectra as a function of the inductor arm length $Y$ in TM and TE polarization (solid lines in Fig. 2(a) left and right panels, respectively), compared with the respective experimental spectra (dashed lines). We can see that the simulations qualitatively reproduce the experimental results.

In TM polarization, the reflectivity spectra show a pronounced dip at low frequency, which red shifts and has a decreasing contrast with increasing inductor length, and a second resonance around 7 THz which is almost independent of $Y$. On the other hand, the TE polarized spectra show a single, red-shifting resonance while the inductor length is increased. In all cases, the simulated resonances are slightly blue-shifted by 0.2–0.3 THz from the experimental ones, and show a better quality factor than experimentally observed. This discrepancy can arise from inhomogeneous broadening of the experimental resonances due to fabrication imperfections, but also from losses in the constitutive materials, especially the SiN layer. [28]

Figure 2(c)-(e) show the most notable features in the electric and magnetic field and/or energy profiles of these resonances, along the planes sketched in Fig. 2(b). Each colormap is normalized to the maximum value in the plot plane.

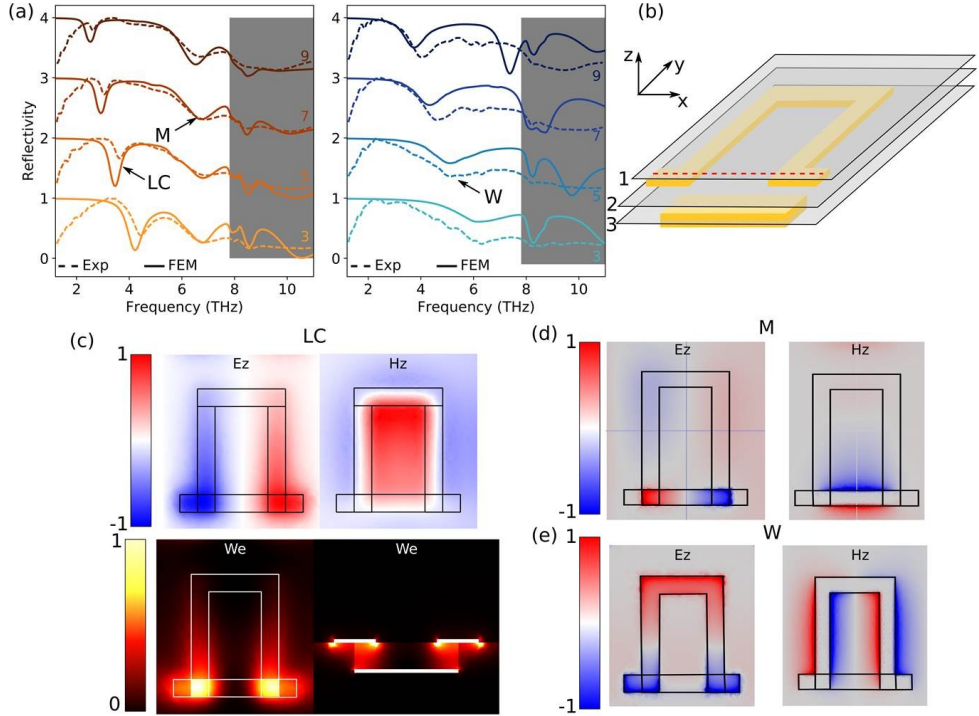

*Figure 2: (a) Simulated (solid lines) and experimental (dashed lines) reflectivity spectra as a function of inductor length Y (indicated on each curve, in µm), for the TM (left) and TE (right) polarizations. (b) Sketch of the cut planes used in the electric and magnetic field maps in panel (c). (c) Top: Vertical component of the electric ($E_z$, left) and magnetic ($H_z$, right) fields in the x-y plane passing through the center of the capacitors (plane 2 in panel (b)). Bottom: Electric energy density (We) for the LC mode in plane 2 (left) and in a x-z plane passing through the middle of the capacitors (right, vertical plane containing the red dashed line in panel (b)). (d): Vertical component of the electric ($E_z$, left) and magnetic ($H_z$, right) fields for the M mode in the plane of the bottom plate (plane 3 in panel (b)). (e): Vertical component of the electric ($E_z$, left) and magnetic ($H_z$, right) fields for the W mode in the plane of the top inductive loop (plane 1 in panel (b)). All colormaps are normalized to the maximum value in the plot plane.*

Let us start by the profiles at the LC mode frequency. In the top panels of Fig. 2(c), we plot the vertical component of the electric ($E_z$) and magnetic ($H_z$) fields energy density in the *x-y* plane in the center of the capacitors (plane 2 in Fig. 2(b)). In the bottom panels we show the electric energy density (We) in the same place (left) and in a vertical plane crossing the center of the capacitors (red dashed line in Fig. 2(b)). This resonance corresponds to the fundamental mode of our meta-atom. We can see that the electric energy is mainly confined inside the capacitive regions, i.e. where the top and bottom metallic plates overlap vertically. The dominant components of the electromagnetic field are the vertical $E_z$ and $H_z$ components, which differ from the usual transverse configuration of an electromagnetic wave. Indeed, the two horizontal parallel plate capacitors impose a vertical electric field, while inductive loop acts as a coil in which the magnetic field is uniform and oriented along the coil axis. We have already studied this mode in greater depth [20], [21], [27], and ascribed it to a circuit-like resonance of the meta-atom, which closely resembles an inductor-capacitor resonance in a LC circuit.

In Fig. 2(d), we plot the vertical component of the electric field ($E_z$, left) and magnetic (Hz, right) energy density in the x-y plane of the lower metallic plate, where it is maximal, at the frequency of the M mode. The vertical component of the electric field (Fig. 2(d)) has maximum amplitude and opposite signs at each end of the bottom plate. The magnetic field circles around the plate indicating a current along the rod axis. These two observations indicate that this mode closely resembles the dipolar (half-wavelength) resonance of the bottom metallic plate. This field configuration explains the fact that the frequency of this resonance, $f_M$=7 THz, is independent of the parameter Y. Indeed, it depends only on the length $L_2$ of the bottom metal strip, $L_2$=5μm, through the formula $f_M = c/(2n_{eff}L_2)$, with c the speed of light and $n_{eff}$ an effective index, for which we find a numerical value $n_{eff}$= 4.2. Such a high value is similar to the one found in patch antennas filled with GaAs [11], meaning that most of the energy of the resonance is located within the GaAs slab.

In Fig. 2(e) we show the vertical component of the electric field ($E_z$, left) and magnetic field ($H_z$, right) for the W resonance. We can see that the electric energy is localized at the two opposite ends of the inductive arms of the meta-atom. As for the M mode, looking at the parity of $E_z$ and $H_z$, we can easily recognize a half-wavelength resonance along the meta-atom length. The corresponding frequency is $f_W = c/[2n_{eff}(Y+4.5\mu m)]$, where Y+4.5 μm is the half length of the upper metallic part including the lateral extensions. With that formula, we find an effective index $n_{eff}$ =2.8, which is above but close to the SiN index, $n_{SiN}$=2.2.

We have thus identified two propagating resonances that are mainly located either at the bottom (M) or the top (W) metallic parts. The LC resonance that involves both metallic patterns is the lowest frequency resonance as expected from a quasi-static approximation [23]. In the following, this resonance is further analyzed in terms of a circuit model.

### 3. Analytical circuit model

We have shown above that numerical simulations can provide insight in the electromagnetic field configurations associated with resonances in the reflectivity spectra. However, in general they fail to reproduce correctly the amplitude of the absorption features. Furthermore, if the fundamental resonance of the structure does operate in the quasi-static limit, it must correspond to an equivalent circuit where the different lumped components depend solely on the geometry of the structure.

Our analytical circuit model is depicted in Figure 3. Our approach to describe the interaction of the periodic array of meta-atoms with free-space radiation is inspired from the methods developed for frequency selective surfaces [29]. As indicated in Fig. 3(a) and 3(b), the free space is treated as a transmission line port of impedance $Z_0$ = 377 Ω and each meta-atom is modeled with an impedance $Z_R$ detailed in Fig. 3(c). The SiN slab shunted by the ground plane is modeled with a frequency-dependent impedance $Z_{SiN}$, which, following Ref. [30], is expressed as: $Z_{SiN}(\omega) = i(Z_0/n_{SiN})\tan(\omega T_{SiN} n_{SiN}/c)$, with $n_{SiN}(\omega)$ the complex refractive index of the SiN layer that uses the same values used in the numerical simulations described above and $T_{SiN}$ is the thickness of the SiN layer. We also consider that meta-atoms are mutually coupled by a capacitance $C_M$ as illustrated in Fig. 3(a). This capacitance is important to describe the effect of the periodic arrangement of the meta-atoms, as previously shown for patch-antenna arrays [24], [31]. In our case, we treat a normally incident field with a constant phase along the array. We can therefore consider that each meta-atom has identical field contributions, such as the ones described in Figure 2. In that configuration the top capacitive pads have opposite charges, and the pads of neighboring meta-atoms can be considered as coupled microstrip lines. The capacitance $C_M$ is therefore the odd-mode capacitance of the fringing fields that connect the two pads, and can be expressed as [24] $C_M = (0.5/\pi)\varepsilon_0 (1+\varepsilon_{SiN}) W \ln\left[ 2(1+\sqrt{k})/(1-\sqrt{k}) \right]$.

Here:

$k = 2\sqrt{2W(S+2W)}/(S+2W)$,

with $S$ = 5 µm the distance between meta-atoms. Using this expression for W=1µm we obtain a numerical value $C_M$ = 10(1+$n_{SiN}(\omega)^2$) aF which has been used for all structures reported here.

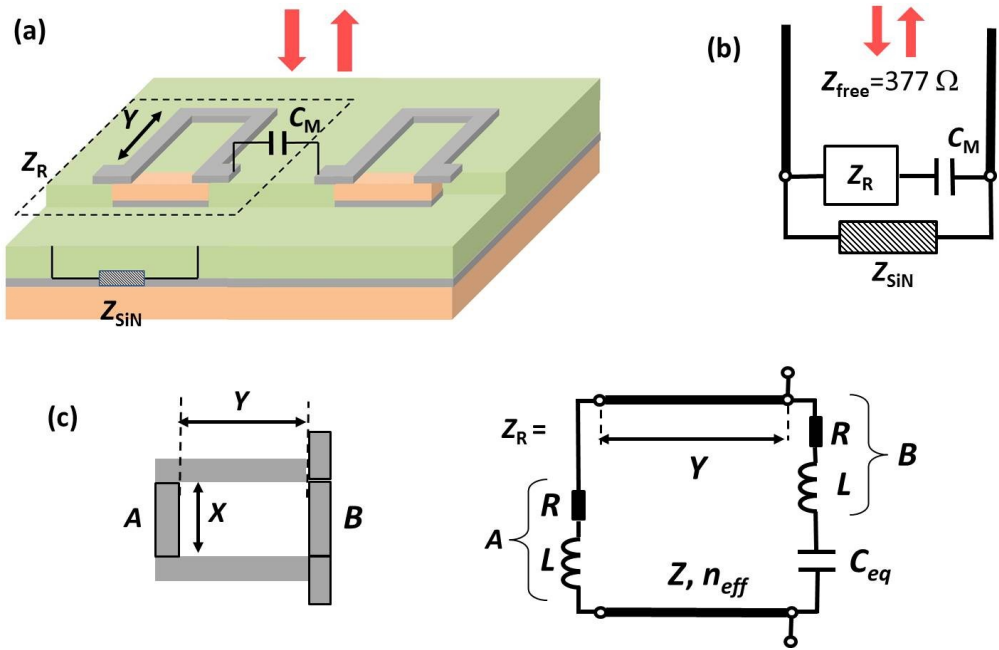

Figure 3: (a) Sketch of the metamaterial showing a unit cell (dashed rectangle) with a transmission line length Y, and a characteristic impedance $Z_R$. The SiN layer is modeled by a resistance $R_{SiN}$, and adjacent unit cells are coupled through a capacitive term $C_M$. (b) Equivalent circuit model showing the arrangement of the unit cell, the capacitor modeling the intercell coupling and the SiN layer. The unit cell coupling to free space is modeled as a transmission line port with free space impedance $Z_{free}$ = 377 Ω. (c) Equivalent circuit model of the resonator. The two segments A and B are modeled with a resistor and inductance in series. The bottom plate is capacitively coupled to the segment B through $C_{eq}$. The radiation resistance is placed where the electric field is confined in the circuit, near the capacitors. The two segments are connected with a transmission line of length Y.

The detailed equivalent circuit for the meta-atom is shown in Fig. 3(c). It is similar to the one reported in Ref. [16], where the parallel section of the meta-atom is treated as a coplanar stripline transmission line of length Y and effective index $n_{eff}$. The impedance of the transmission line is then Z = $Z_0$/(1.14 $n_{eff}$), where the factor 1.14 takes into account the geometry of the metal strips [25]. In order to simplify our analysis, the transmission line is considered to be lossless. The impedance of the two metal segments A and B of length X = 3 µm shown in Fig. 3(c) are modeled as an inductance L in series with resistor R that describes the ohmic loss in the system. The inductance is modeled following Eq.(2) in Ref. [27]. The segment B (lower ground plate) is capacitively-coupled with the rest of the structure through the capacitances of the double-metal regions; as these capacitances must be considered in series, we describe them as a single capacitance $C_{eq}$. The latter can be expressed as

$$C_{eq} = 0.5\varepsilon_{GaAs}\varepsilon_0 W^2 / T_{GaAs} + \Delta C,$$

where the first contribution is a parallel plate capacitance and the second $\Delta C$ takes into account the fringing field corrections as described in Ref. [21]; in the present case we obtain $C_{eq}$ = 74 aF for all structures.

The values of the reflectivity minima in the spectra are dependent on the balance between the coupling with the incoming radiation and the dissipation of the system [32]. In our model, the coupling with an incident radiation is related to the fringing fields [27], which are modeled as a lumped capacitor $C_M$, while the dissipation in the circuit is quantified by the resistance $R$. To model the resistance $R$ adequately we need to take into account high frequency effects such as the skin effect and current crowding [33]. The high frequency resistance can then be expressed as $R = R_{DC}F(f)(1+(2\pi f\tau)^2)$, where $R_{DC}$ is the expected DC value of the resistance and the function $F(f)$ depends on frequency and the geometry of the resistor and has been explicitly provided in Ref. [33]. In our case, the metal strips are $X$ = 3 μm long, $W$ = 1 μm wide and have a thickness $t_{Au}$ = 0.15 μm (the thickness of the gold layer deposited by metal evaporation). Using the expression of the skin depth $\delta$= 0.075 [μmTHz$^2$]/f$^2$ [27] we obtain the following expression $F(f)$ = $(1.48-0.48\exp(-0.2f^{0.5}))/(1-\exp(-1.16/f^{0.5}))$, where $f$ is the frequency expressed in THz. The factor $(1+(2\pi f\tau)^2)$ arises from the Drude model [34], with $\tau$ the scattering time, which is discussed further below.

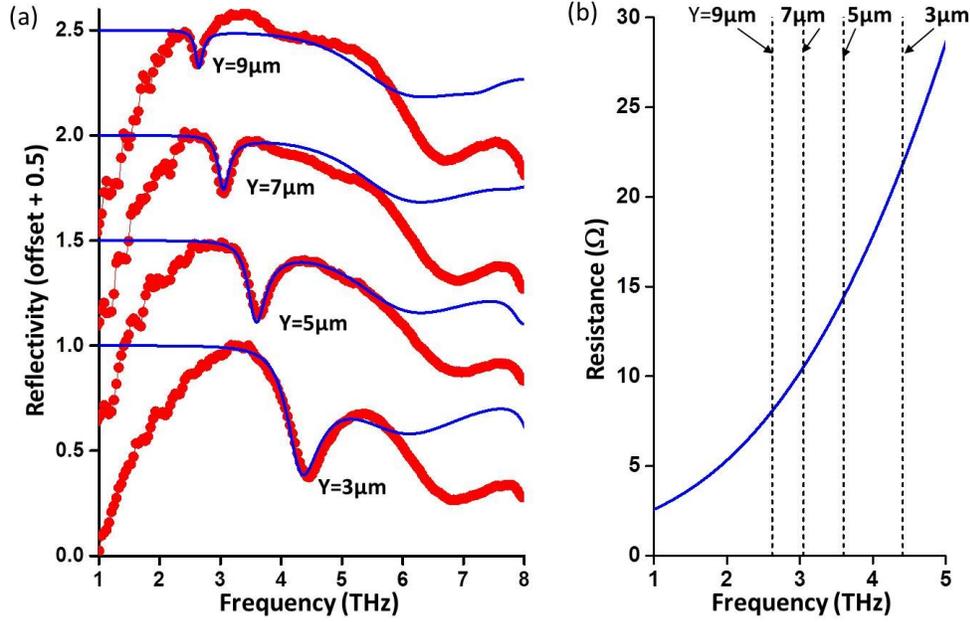

Figure 4: (a) Simulated reflectivity obtained from the circuit model (blue lines) compared with the experimental data (red circles) for the different values of the arm length Y. (b) Resistance R of the inductive segments A and B (see Fig. 3) as a function of frequency.

Having established the parameters of our circuit, we compare our model with experiments. The comparison is shown in Fig. 4(a). The reflectivity is computed from the full circuit in Fig. 3(b) as the ratio between the reflected and incoming power in the transmission line of impendence $Z_0$. The latter can also be expressed as $r(f) = |(Z_0-Z_{tot})/(Z_0+Z_{tot})|^2$, where $Z_{tot}(f)$ is the total impedance terminating the transmission line. The reflectivity curve is adjusted to the data using three fit parameters $\tau$, $R_{DC}$ and $n_{eff}$. The parameter $n_{eff}$, the effective index of the transmission line of the meta-atom, influences only the resonant frequency of the reflectivity dip. Very good agreement of the frequency of the reflectivity resonances is obtained using a single value $n_{eff}$ = 1.92 for all structures. This value is close to the refractive index of the SiN layer ($n$ = 2.2), but lower owing to leakage of the fields into the air region. With all circuit parameters fixed, the width and depth of the reflectivity features are mainly governed by the quantities $\tau$ and $R_{DC}$. Very good agreement with the shape of the reflectivity dips for the *LC* resonance are obtained using the values $R_{DC}$ = 1.2 $\Omega$ and $\tau$ = 0.08 ps. The corresponding frequency-dependent resistance $R$ is plotted in Fig. 4(b). The dashed vertical lines in the plot indicate the position of the resonance frequency for each of the four structures. The obtained value $R_{DC}$ = 1.2 $\Omega$ is twice the one expected from the formula $R_{DC} = \rho X/Wt_{Au}$ = 0.5 $\Omega$ where $\rho$ = 2.44x10$^{-8}$ $\Omega$m is the resistivity of gold. The discrepancy can be ascribed to the granular morphology of the thin film gold layers obtained by metal evaporation, that tend to increase the resistivity [35]. Our overestimated value could also arise from neglecting the resistive losses of the transmission line segments. The value $\tau$ = 0.08 ps of the scattering time in the Drude model is within the range of values (0.055ps -0.22ps) reported by various authors.[36]–[38]

Our circuit model therefore reproduces very well the position of the LC resonances together with their linewidth and reflectivity contrast. We now exploit this model in order to quantify the electric energy distribution in the different parts of the resonator. Ideally, such structures were designed with the hope that all of the electric energy is concentrated in the capacitive parts described by the equivalent capacitor $C_{eq}$. However, the transmission line section will lead to leakage of the electric energy, as already seen in the field maps of Fig. 2. Our model allows this effect to be quantified for the various structures, as shown in Fig. 5.

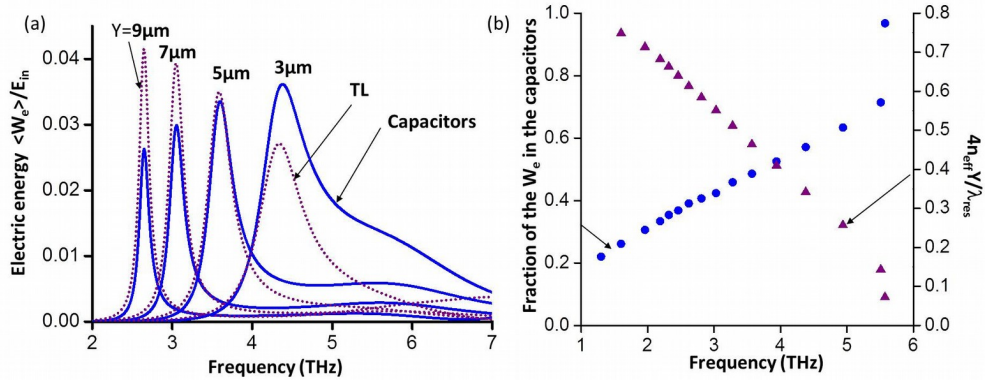

Figure 5: (a) Electric energy localized inside the capacitors (blue solid lines) and in the transmission lines (purple dashed lines) extracted from the circuit model calculations for the four different geometries. (b) Left axis (blue circles): fraction of peak electric energy stored inside the capacitors as a function of the peak frequency. Right axis (purple triangles): Degree of subwavelength confinement.

In Figure 5(a) we plot the electric energy stored in the lumped capacitors as a function the frequency (blue solid lines), in comparison with the electric energy stored in the transmission line (dotted curves). In order to have dimensionless quantities, the electric energy has been normalized to the quantity $P_{in} \times T = P_{in} \times (2\pi/\omega)$, with $P_{in}$ the incoming power from the free space; therefore $P_{in} \times T$ is the energy incoming into the system during a period of oscillation. Both energies display a resonant behavior, and the *LC* mode appears to be the sole mode that leads to a significant electrical energy build-up in that spectral region. We see that generally the electric energy in the transmission line is larger than the energy stored in the capacitors. In Fig. 5(b), we plot the fraction of the peak electric energy in the capacitors as a function of the resonant frequency of the meta-atom. For this plot we have varied the transmission line length *Y* from 0.5µm to 24µm. In the second y-axis of Fig. 5(b) we also plot the quantity $4n_{eff}Y/\lambda_{res}$ which is a measure of the degree of subwavelength confinement in the structure. Indeed, it is the resonant wavelength of the lowest propagating mode in a transmission line with an open and shorted end. In the geometries explored, only 30% to 50% of the total energy is contained in the capacitors, and only for very short lines (Y=0.5µm) is the majority of the electric energy localized in the capacitors. This study shows that the transmission line section is responsible for strong delocalization of the electrical energy away from the double-metal capacitive regions, and care must be taken in order to avoid this effect for an efficient coupling with the electronic transitions in the capacitive regions.

## 4. Optimization of the electric field confinement

The purpose of these structures is to provide a strong confinement of the electric energy inside extremely small semiconductor volumes defined by the capacitive parts of the meta-atoms. Both the FEM simulations and the circuit model bring physical information on the way to optimize this behavior. The circuit model revealed the electric field leakage along the inductive part of the resonators. Another effect that reduces the field confinement is the shape of the capacitors themselves. Indeed, optimal field confinement can be achieved in a perfect parallel plate capacitor, where the field is localized solely inside the capacitor plates. When the typical width *W* of the plates becomes comparable to the thickness of the spacer material *d*, then a significant portion of the electric energy is located in the capacitor fringing fields [39], as sketched in the inset of Fig. 6. This effect is already seen from the FEM simulations in Fig. 2(c).

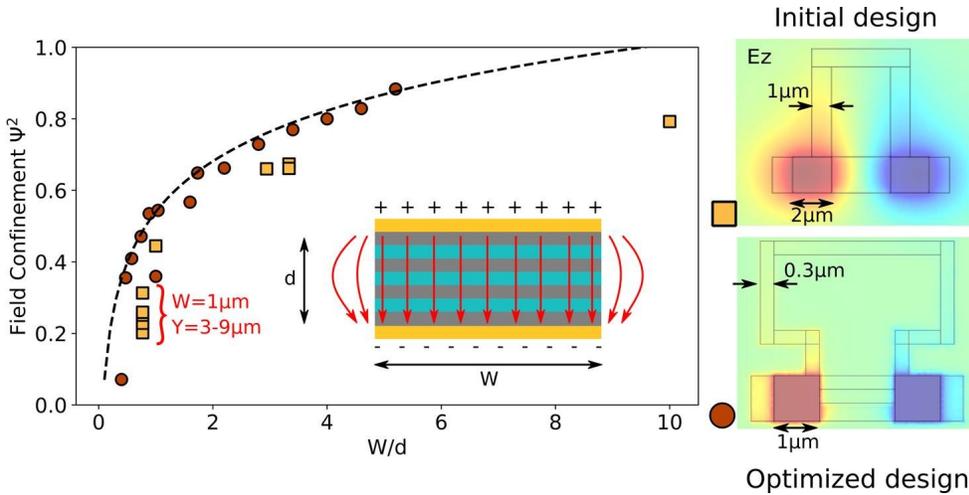

Figure 6: Effect of the aspect ratio W/d of the capacitive parts of the meta-atom on the electric field confinement. The results are obtained from FEM simulations by integrating the electric energy stored inside the capacitor volume, normalized by the total electric energy. The black dashed line is a guide for the eye following a logarithmic trend. Two geometries are considered, as shown on the right: straight transmission lines (yellow squares), and broken transmission lines (brown circles). The simulation results show the vertical component of the electric field $E_z$.

In order to optimize the electric field confinement, we implemented a new design that solves the aforementioned issues (Figure 6 lower left panel). To decrease the influence of the transmission line, it has been broken into shorter sections. Furthermore, the transmission line has smaller width (300nm) than the capacitor (1µm), in order to increase the impedance mismatch between the two.

We have further investigated the influence of the capacitor thickness on both designs. For this we define a field confinement factor $\Psi^2 = \int_{GaAs} \varepsilon_{GaAs} |E_z|^2 / \int_{\infty} W_e^2$, where we integrate the electric energy density in the vertical polarization of the electric field inside the GaAs active region and normalize it by the entire electric energy density in the resonator mode. We present in Fig. 6 the value of the field confinement factor for different structures, changing both the aspect ratio and the geometry of the inductive loop.

Yellow squares refer to the straight and parallel inductor arms geometry as presented above, while brown circles represent values obtained for an inductor with sharp bends, as shown in the bottom right panel. The four initial structures discussed in Fig. 1 and 2 are indicated (where W = 1 µm and Y ranges from 3 µm to 9 µm). For both geometries we varied both the capacitor width (W) and the thickness (d), and present the results as a function of the aspect ratio (W/d). In the case of the initial design, two capacitor widths were investigated (W = 1 µm and 2.6 µm), varying the aspect ratio by simulating structures of different thicknesses (d). In the case of the optimized design, two sets of simulations were performed: one had a fixed capacitor width (W = 2.6 µm) and the capacitor thickness was varied (d), the other using a fixed thickness (d = 500 nm) and the capacitor width (W) was varied. The black dashed line is a guide for the eye following a logarithmic trend, ~log(W/d). Looking first at the results from the straight inductor geometry (yellow squares), increasing the aspect ratio improves the field confinement. For a given aspect ratio, fabricating inductive loop wires thinner than the capacitors induces an impedance mismatch, limiting the electric field leakage outside the capacitor and hence increasing the field confinement. From the graph of Fig. 6 we see indeed that, in the case of the optimized design (brown circles), aspect ratios $W/d$ >3 are already suffiient to achieve 80% confinement. This is illustrated in the field maps on the lower left panel of Fig. 6. Combining such aspect ratios with bent inductor lines thus seems to be the best strategy to achieve very small electric field mode volumes for a given THz frequency.

The substantial improvement of the resonator geometry can be demonstrated by studying the coupling between the LC mode of the structure and an electronic excitation sustained by the two-dimensional electron gas inserted inside the semiconductor region in the capacitors. If the overlap between the circuit mode and the electronic excitation of the quantum well is large enough, the system enters the strong coupling regime, where energy is reversibly exchanged between the electromagnetic mode and the intersubband transitions in the QWs. As a result, the LC resonance observed in reflectivity experiments is split into two polariton states. If the LC resonator and the intersubband absorption are perfectly matched, then the frequency splitting between the two polariton states is $2\Omega_R \propto \sqrt{\Psi^2}$, where $\Psi^2$ is the confinement factor of the z component of the electric field $E_z$, which is the only component coupled to the intersubband absorption. The polariton splitting observed in reflectivity spectra can thus be used as a direct probe of the electromagnetic confinement of the resonator, as already shown in Ref. [14], [20]. The splitting $2\Omega_R$, also called the vacuum Rabi splitting, can be determined precisely by changing the circuit resonance: one thus obtains the dispersion of the polariton states, and $2\hbar\Omega_R$ is the minimum energy difference between the high energy (UP) and low energy (LP) polaritons.

In Figure 7 we present the optical spectroscopy of two sets of resonators. The first one (Fig. 7(a),(b)) is the set that we presented so far, where the capacitor has an area of 1 x 1 µm² while the active region contains 20 doped QWs, resulting in a thickness of 1.3 µm, and hence an aspect ratio of 0.77 for the capacitors. Looking at Fig. 6, we thus expect a low field confinement for this structure (<0.4). The second set corresponds to resonators with bent inductors as detailed in the last paragraph, and with an active region containing only 5 QWs. Keeping the capacitor width constant, this results in a capacitor aspect ratio of 3.3, and hence a confinement factor of ~0.8 according to Fig. 6. These studies have been performed at liquid helium temperature where the quantum well transitions become active [20].

As it can be seen in Fig. 7 (a),(b), the first version of the LC resonators could not enter the strong coupling regime. We show in Fig. 7(a) a comparison between the reflectivity of the metamaterial at room (blue line) and low (red line) temperature. We can see that the circuit resonance shifts in frequency when cooling the sample. However, as can be seen from Fig. 7(b), the dispersion relation of the intersubband absorption and circuit mode shows a crossing, evidencing the fact that the system is in the weak coupling regime, due to a low field confinement. In Fig. 7(c) we show the reflectivity of the optimized metamaterial. Here, the resonator mode splits into two resonances (lower and upper polaritons LP and UP respectively). The dispersion relation presented in Fig. 7(d) shows the anticrossing of the LP and UP modes, with a vacuum Rabi splitting of 0.7 THz. Following the experimental procedure of Ref. [20], we can derive the value of the electric field confinement from the vacuum Rabi splitting. We find a value $\Psi^2 \sim 0.8$, in good agreement with the FEM simulations.

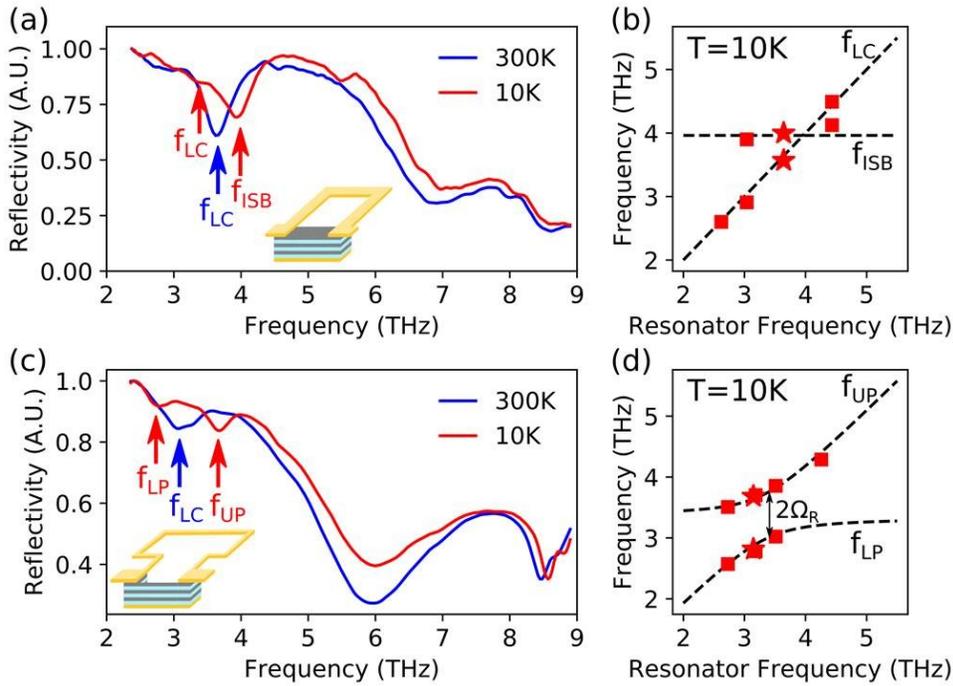

Figure 7: (a) Reflectivity spectra of the LC metamaterial at room temperature (blue line) and at 10K (red line). The arrows represent the resonant frequencies of the LC resonator and the intersubband transition. (b) Low temperature dispersion relation of the modes showing the crossing between the intersubband and the LC resonance. The stars mark the device for which the results are shown in (a). (c) Reflectivity spectra of the optimized LC metamaterial at room temperature (blue line) and low temperature (red line), showing the frequency splitting of the intersubband and LC resonances into the upper and lower polariton modes. (d) Dispersion relation of the upper and lower polaritons showing the anticrossing and the characteristic vacuum Rabi splitting $2\Omega_R$.

## 5. Conclusion

In conclusion, we have presented an extensive discussion on the different modes sustained by our three-dimensional LC meta-atom. Starting from an initial geometry of the meta-atom, we complement optical reflectivity experiments with FEM simulations revealing the different nature of the resonances observed in the reflectivity spectra. We proposed a circuit model of the meta-atom array, showing a good agreement with the experimental data and numerical simulations, allowing us to gain insight into the radiative and non-radiative properties of the meta-atoms. Furthermore, the circuit model also provides new insight on the electric energy localization in the capacitors of the LC meta-atom. This is a key parameter in the future use of these meta-atoms in optoelectronic devices in which a semiconductor active region is inserted inside the meta-atom capacitors. Using this new insight, we proposed an optimized design of the LC meta-atom ensuring an extreme confinement of the electric field inside the capacitors [20]. A clear advantage of such an approach is to identify the different lumped elements that comprise the resonator, and to ascribe their capacitance, inductance and resistance solely from their geometrical characteristics. This approach, which is common in the microwave region, is more difficult to implement in the THz domain, where the displacement current usually has an important contribution. This approach can also be very useful when considering integrating active elements, such us semiconductor or superconductor devices.

**Funding**

This work was supported by the French National Research Agency under the contract ANR-16-CE24-0020, and the Engineering and Physical Sciences Research Council (UK) grant EP/P021859/1. EHL acknowledges the support of the Royal Society and Wolfson Foundation.

**Acknowledgments**

The authors acknowledge the help of the technical staff from the cleanroom facility of Université Paris Diderot.

**Disclosure**

The authors declare no conflicts of interest.